\documentclass[usegraphics]{emulateapj}

\shorttitle{A kinematic search for globular cluster tidal tails}
\shortauthors{Kiss et al.}

\begin{document}

\title{A wide-field kinematic survey for tidal tails around five 
globular clusters}

\author{L. L. Kiss\altaffilmark{1}, P. Sz\'ekely\altaffilmark{2}, 
T. R. Bedding\altaffilmark{1}, G. \'A. Bakos\altaffilmark{3}, G. F. Lewis\altaffilmark{1}
}
\altaffiltext{1}{School of Physics, University of Sydney, NSW 2006, Australia;} 
\altaffiltext{2}{Department of Experimental Physics, University of Szeged, 
D\'om t\'er 9. Szeged 6720, Hungary}
\altaffiltext{3}{Harvard-Smithsonian Center for Astrophysics, Cambridge, 
MA, USA}

\begin{abstract}

Using the AAOmega instrument of the Anglo-Australian Telescope, we have obtained
medium-resolution near-infrared spectra of 10,500 stars in two-degree fields centered on
the galactic globular clusters 47~Tuc, NGC~288, M12, M30 and M55.  Radial velocities and
equivalent widths of the infrared Ca II triplet lines have been  determined to constrain
cluster membership, which in turn has been used to study the angular extent of the
clusters. From the analysis of 140--1000 member stars in each cluster, we do not find
extended structures that go beyond the tidal radii. For three cluster we estimate a 
1\% upper limit of extra-tidal red giant branch stars. We detect systemic rotation 
in 47~Tuc and M55.   

\end{abstract}


\section{Introduction}

The present structure of our Galaxy is result of a complex dynamical evolution in space
and time. Globular clusters (GCs) are excellent tracers of this evolution because they
interact with the gravitational field of the Milky Way, which affects their internal
dynamics (Meylan \& Heggie 1997). When passing through the galactic  disk or the bulge,
tidal shocks occur and these can have dramatic effects on the cluster evolution (Gnedin et
al. 1999, Dehnen et al. 2004). N-body simulations have shown that the resulting tidal
tails may be used to trace the orbital paths of GCs, which in turn reveals information
about the enclosed mass of the Galaxy (e.g., Combes et al.\ 1999; Capuzzo Dolcetta et al.\ 2005).

Recent star count surveys have revealed tidal tails stretching many degrees beyond
the cluster tidal radius, for which the best cases include Palomar~5 and NGC~5466 
(Odenkirchen et al. 2001, 2003; Belokurov et al. 2006; Grillmair \& Johnson 2006). 
These tails are believed to be cluster stars that  have escaped due to tidal
shocking.  A crucial ingredient in these
discoveries was the statistics of stars: in order  to detect a non-circular
over-density
around a cluster, one has to eliminate galactic field stars very
carefully. Escaped cluster members are usually identified by their colours and apparent
magnitudes, using optimal contrast filtering in the color-magnitude plane.

This approach, however effective in distant and unreddened clusters, cannot easily be used
for nearby clusters that have strong galactic star  contamination  and/or heavy
interstellar reddening. However, these clusters could play a very important role in
constraining models of tidal disruption, because they represent the state of a GC
immediately before/after a gravitational shock by passing through the disk. The recently
commissioned AAOmega fiber-fed multi-object spectrograph (Sharp et al. 2006) on the 3.9m
Anglo-Australian Telescope offers a unique possibility for measuring radial velocities
with 1--2 km/s precision for up to 350--360 stars per exposure in two-degree field of
view, which enables a spectroscopic distinction between field stars and GC members over
several tidal radii. In this Letter we report on the first
results of a such survey for five southern GCs.

\section{Observations}

The targets (Table\ \ref{sample}) have been selected from the catalogue of Harris
(1996), using the following criteria: {\it (i)} nearby, thus with low or moderate
reddening; {\it (ii)} small tidal radius ($r_{\rm t}\lesssim$20$^\prime$), so that  the
two-degree field of the instrument covers an area that is several times $r_{\rm t}$
across; {\it (iii)} large radial velocity compared to the galactic field, which allows
membership identification. With telescope time granted in
August 2006, the two best GCs were M55 and M30, both with very large radial
velocities. We also consulted the tidal tail study of Leon et al. (2000), which
resulted in adding  NGC~288 to the targets. M12 was included because of the
suggested severe tidal stripping (de Marchi et al. 2006), while 47~Tuc for its known
systemic rotation (Meylan \& Mayor 1986). Two of these objects, M55 and M12, lie
outside the ``surviving'' boundaries of GCs in figs. 21-24 of 
Gnedin \& Ostriker (1997), favouring the possibility of significant tidal tails.  

\begin{figure*}
\begin{center}
\leavevmode
\includegraphics[height=7cm]{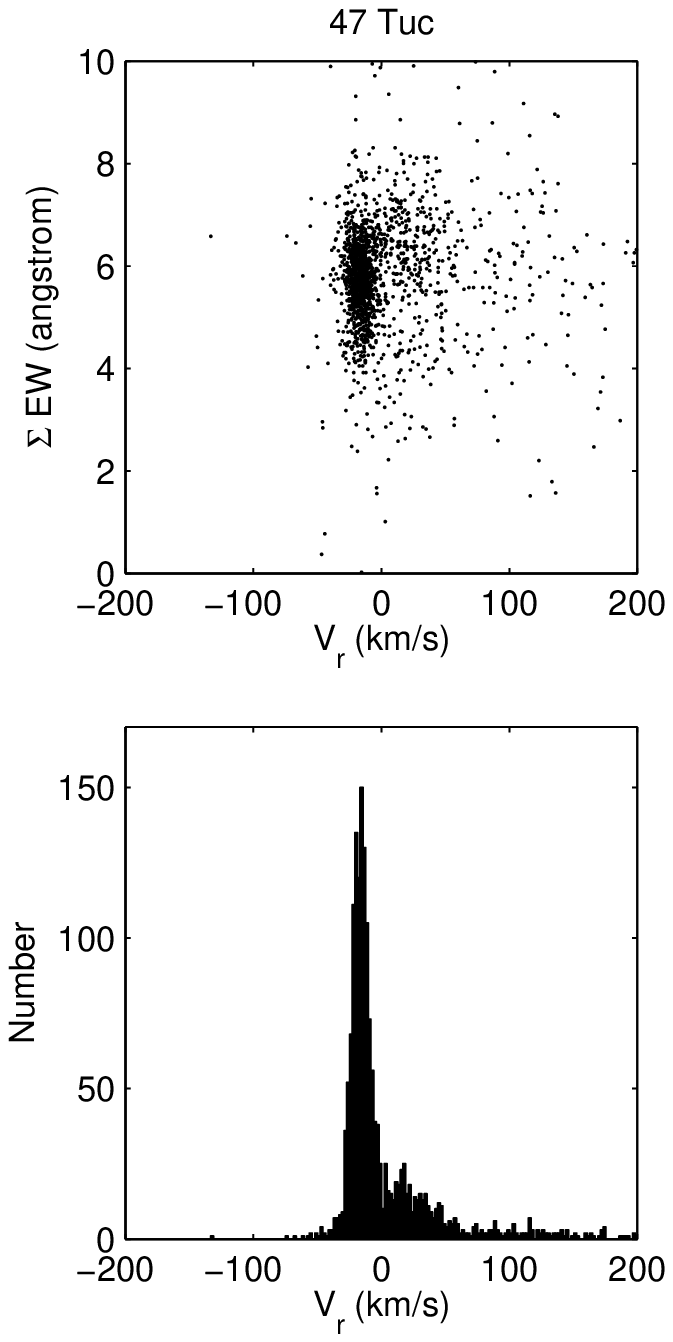}
\includegraphics[height=7cm]{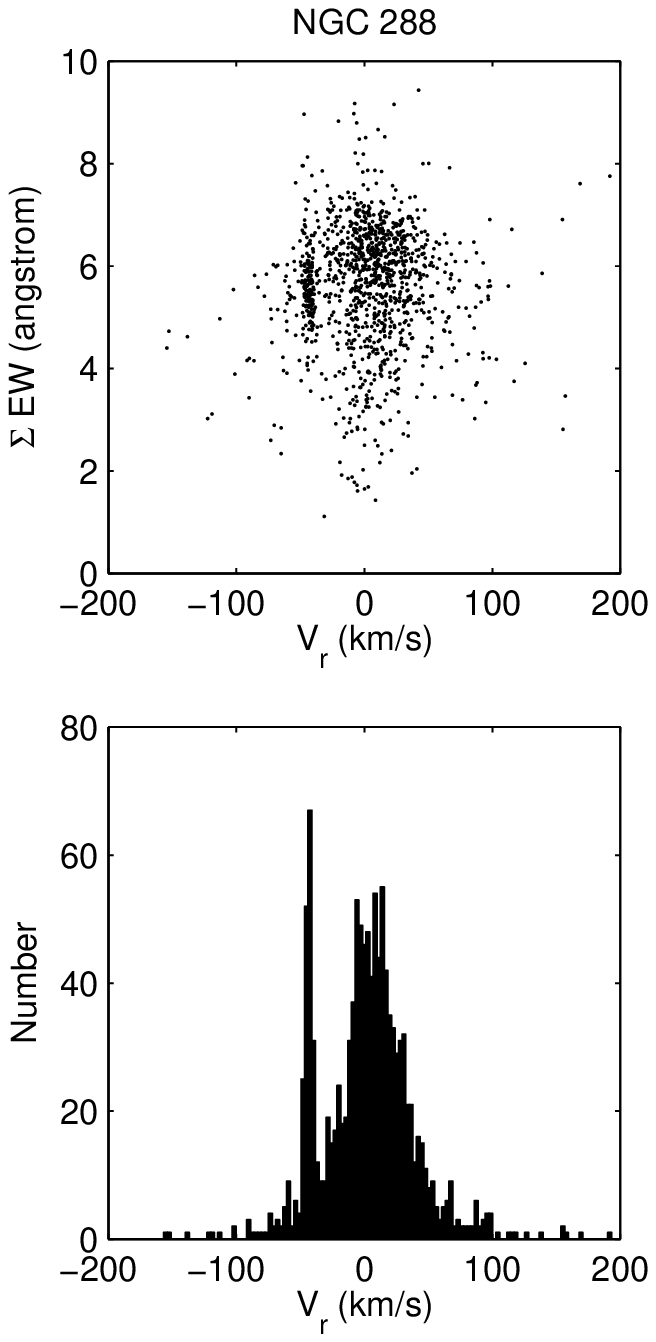}
\includegraphics[height=7cm]{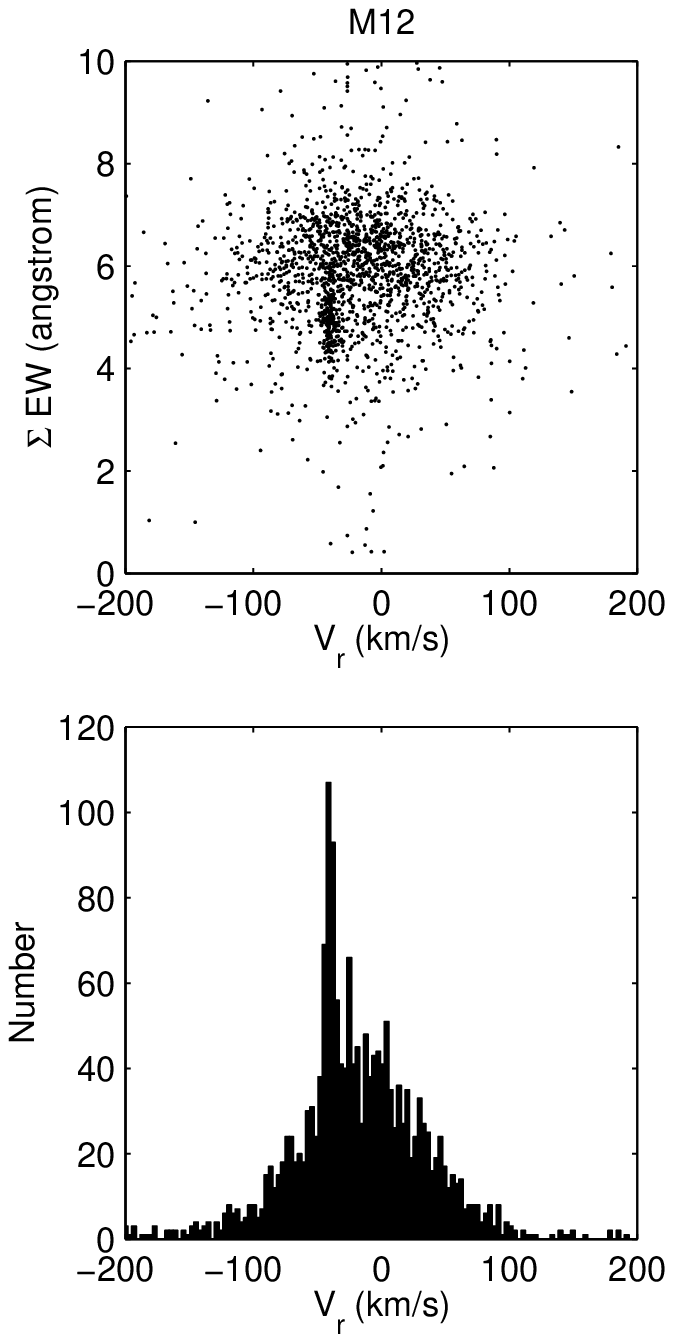}
\includegraphics[height=7cm]{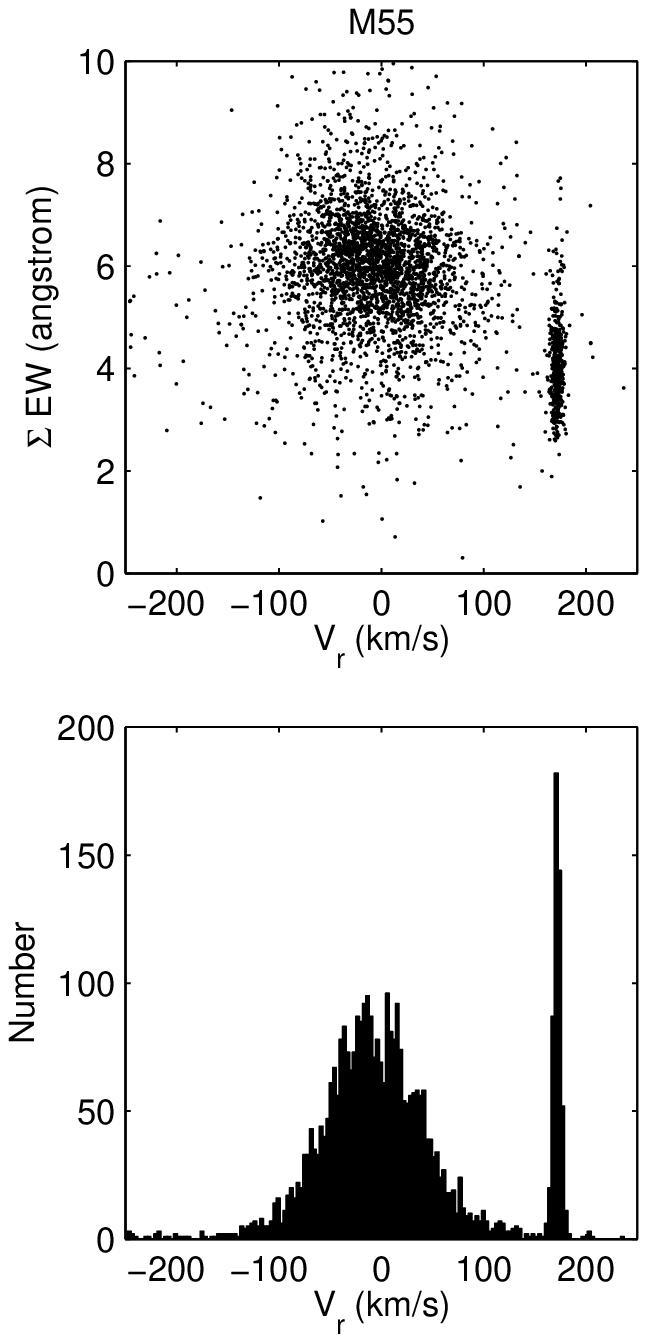}
\includegraphics[height=7cm]{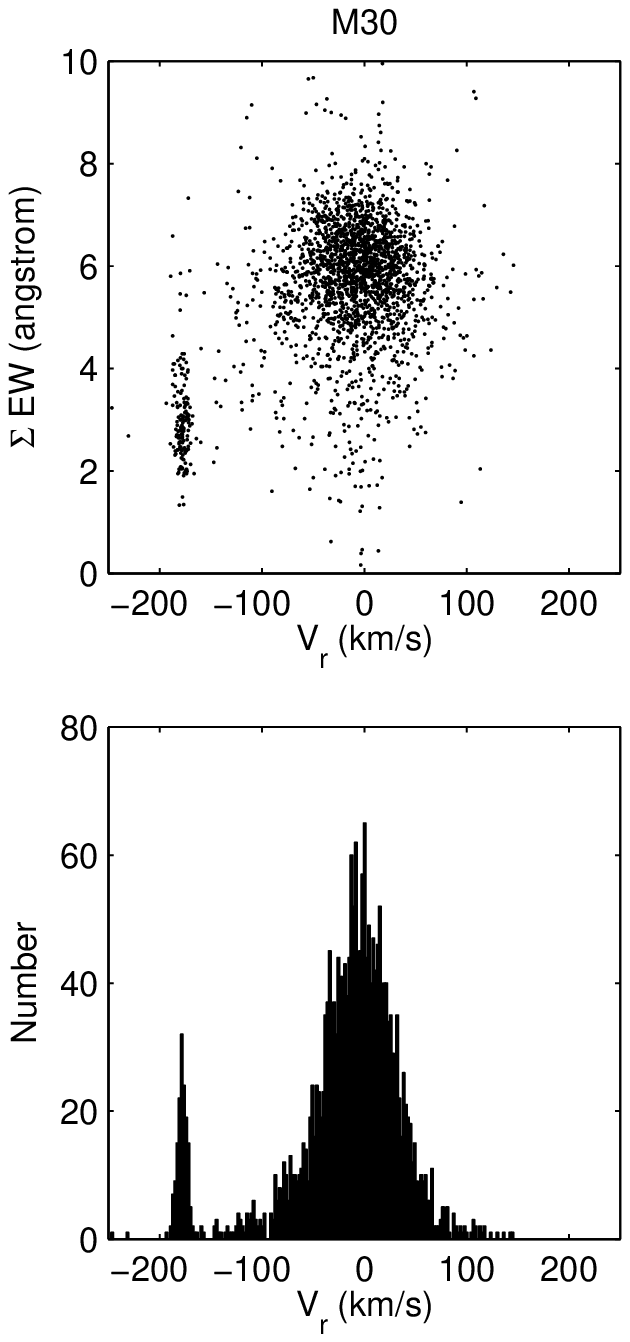}
\caption[]{\label{members} {\it Upper panels:} Radial velocities and total 
equivalent widths of the Ca II triplet for the observed stars. {\it Lower panels:}
Radial velocity histograms that were used to determine the velocity ranges of the
clusters. Metallicity decreases from left to right.} 
\end{center}
\end{figure*}

The observations were carried out on six and a half clear nights, between  August 12 and
August 18, 2006, with typical seeing between 1.2$^{\prime\prime}$ and
2.2$^{\prime\prime}$. The data were taken with the D1700 grating that has been optimized
for recording the Ca~II infrared triplet region. The spectra range between 8350
\AA\ and 8790 \AA, with a resolving power of  $\lambda/\Delta\lambda=$10,500. Each
cluster was observed with several field configurations, which were all centered on the
cluster. For every configuration we exposed 60 to 90 minutes, giving signal-to-noise
ratios 50 to 250.

Of the 392 fibers of the instrument, we used 30 for sky background measurements
and 5 to 8 for guiding. Lists of target stars were prepared using the  2MASS
point source catalog (Skrutskie et al. 2006). For M55, we selected stars that matched  the
color and magnitude range of the red giant branch. For M30 and NGC~288, we
did not restrict the sample, while for 47~Tuc and M12 we selected the faint end of the red
giant branch. The $K$ magnitudes of the selected stars fell between 8 mag and  15 mag. To
minimize cross-talk of the fibers due to scattered light, each field configuration was
limited to stars in a 3-mag wide range. For this, $V$- and $I$-band magnitudes were
estimated from the 2MASS $JHK$ magnitudes, using a set of linear transformations
${B,V,R,I}= a_X J+b_X H+c_X K_S+d_X$, which was calibrated using  equatorial standard
stars from Landolt (1983, 1992), cross-identified with the 2MASS catalogue ($X$ refers to
the Johnson bands).  These transformations gave a useful and robust relation between
Landolt and 2MASS, with the r.m.s. of the fits about 0.1 mag or less. 

The spectra were reduced using the standard {\tt 2dF} data reduction pipeline that
extracts the wavelength calibrated spectra automatically. We performed continuum
normalization separately and cleaned the strongest skylines
that had residuals left, using linear interpolation of the surrounding continuum. 

\begin{table}
\begin{center}
\caption{\label{sample} The observed globular clusters and their parameters.}
\begin{tabular}{lclrcl}
\hline
Cluster   & d     & $[Fe/H]$ & $V_{\rm r}$~~ & $r_{\rm tidal}$ & $c$ \\
          & (kpc)           &          &  (km/s) &  ($^\prime$)\\  
\hline
NGC 104 (47 Tuc) & 4.5 & $-$0.76 & $-$18.7 & 42.9 & 2.03\\
NGC 288 & 8.8 & $-$1.24 & $-$46.6 & 12.9& 0.96\\
NGC 6218 (M12)   & 4.9   & $-$1.48 & $-$42.2 & 17.6& 1.39\\
NGC 6809 (M55) & 5.3 &  $-$1.81 & 174.8 & 16.3 & 0.76\\
NGC 7099 (M30) & 8.0 &  $-$2.12 & $-$181.9 & 18.3& 2.50c\\
\hline
\end{tabular}
\end{center}
\end{table}

In total, we collected spectra for 10,500 stars, from which we determined 
two parameters for this paper. One is the equivalent width of the
Ca~II triplet lines at 8498, 8542 and 8662 \AA. Following Cole et al.
(2004), we fitted a sum of a Gaussian and a Lorentzian function to each line, which was
then integrated to calculate the equivalent widths and their sum $\Sigma {\rm
EW}=W_{8498}+W_{8542}+W_{8662}$. This parameter is sensitive to surface gravity and
metallicity, which can help constrain cluster membership. The fit also gave an initial
radial velocity (RV) value, which was improved as follows. Using the extensive synthetic
spectrum library of Munari et al. (2005), we selected best-fit synthetic spectra,
degraded in resolution to match the observations, with $\chi^2$ fitting, treating
RV as a fixed parameter given by the line-profile fitting. Each observed
spectrum was then cross-correlated with the best-fit model, resulting in an improved
RV that was usually within $\pm$5 km/s of the initial value and is believed to be
accurate within $\pm$1-2 km/s. 

\section{Membership determination}

We plot $\Sigma {\rm EW}$ vs. RV in the upper panels of  Fig.\ \ref{members}. In each
panel the narrow clump corresponds to cluster members and the broader distribution
comprises field stars. The clusters are ordered by  decreasing metallicity from left to
right and it is also  noticeable how the mean equivalent width decreases with decreasing
metallicity. 

To measure the velocity ranges of the clusters, we analysed the velocity  histograms shown
in Fig.\ \ref{members}. For M55 and M30, the narrow peaks of the histograms give very
well-defined velocity limits. For the other three clusters we determined the velocity
positions of the sudden rises in the histogram and  applied an edge-detection algorithm
that is searching for the extrema of the first derivative. The results are listed in the
2nd and 3rd columns of Table\ \ref{stats} (with uncertainties of about 1 km/s) and their
straight averages agree within 1-2 km/s with the catalogued GC RVs (Table\ \ref{sample}).
We also experimented with fitting a sum of a Gaussian and a low-order polynomial
background to measure the velocity centroid and dispersion of the clusters but found that
Gaussians provided poor fits. 

Using the determined velocity limits, we defined the core samples of candidate members,
i.e. stars with $V_{\rm r}^{\rm min}\leq V_{\rm r} \leq V_{\rm r}^{\rm max}$. However, it
is obvious from the upper panels in Fig.\ \ref{members} that even for M55 and M30, there
is contamination from the field. We estimate the extent of this contamination with the
following simple analysis.

\begin{figure}
\begin{center}
\leavevmode
\includegraphics[width=7cm]{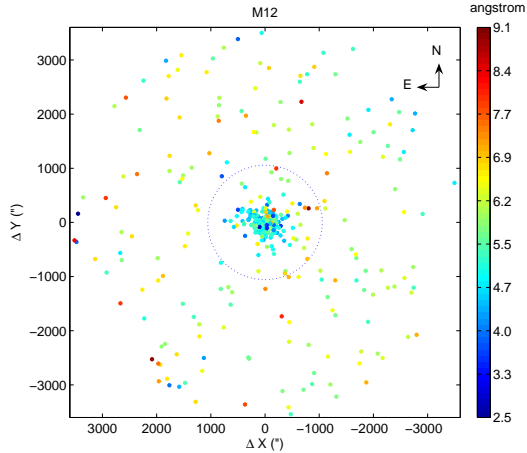}
\caption[]{\label{m12} Positions of stars within the velocity 
range of M12, color coded by $\Sigma EW$. The dotted circle shows the tidal radius.} 
\end{center}
\end{figure}

In Table\ \ref{stats}, $N_{\rm tot.}$ represents the total number of observed stars in
each field, while the core samples contain $N_{V_{\rm r}}$ objects. The question is how
many of them have only by chance the same velocity as the cluster. To answer this, we
removed the cluster peak from the histogram and fitted a smooth, low-order polynomial
over $V_{\rm r}^{\rm cluster}\pm$50 km/s. Summing up the  fitted function through the
cluster peak gave us an estimated number $N_{\rm hist.}$  of contaminating field stars
with an uncertainty of about 6-8\%.
We also counted the number of stars in the core samples that are located
outside the tidal radius ($N_{r_{\rm t}}$ in Table\ \ref{stats}). This number
was multiplied by ($60^2/(60^2-r_{\rm t}^2)$) to take the missing area inside the tidal
radius into account (60$^\prime$ is the field radius). The resulting number,  
$N^\prime_{r_{\rm t}}$, gives another estimate of the field stars in the core sample.

\begin{table}
\begin{center}
\caption{\label{stats} Statistics of stars in the 2-degree fields of view.
See text for the abbreviations.}
\begin{tabular}{lrrrrrrr}
\hline
Cluster          & $V_{\rm r}^{\rm min}$ & $V_{\rm r}^{\rm max}$& $N_{\rm tot.}$ 
& $N_{V_{\rm r}}$ & 
$N_{\rm hist.}$ & $N_{r_{\rm t}}$ & $N^\prime_{r_{\rm t}}$\\
\hline
47~Tuc & $-$32 & 0 & 1696 & 1149 & 171 & 70 & 140 \\
NGC~288 & $-$51 & $-$38 &  1219 & 179 & 30 & 33 & 35\\
M12 & $-$52 & $-$31 & 1780 & 407 & 200 & 175 & 192\\
M55 & 160 & 182 & 3571 & 501 & 8 & 7 & 8 \\ 
M30 & $-$195 & $-$165 & 2240 & 154 & 7 & 4 & 5\\
\hline
\end{tabular}
\end{center}
\end{table}

The good agreement between $N_{\rm hist.}$ and $N^\prime_{r_{\rm t}}$ in most cases
suggests that stars outside the tidal radii belong to the galactic field only. The
largest difference was found for 47~Tuc, but its tidal radius covers over 2/3 of the
whole field, so that the few stars outside do not  provide an accurate estimate of the
total contamination. 

\section{Celestial distribution of member stars}

\begin{figure*}
\begin{center}
\leavevmode
\includegraphics[width=7cm]{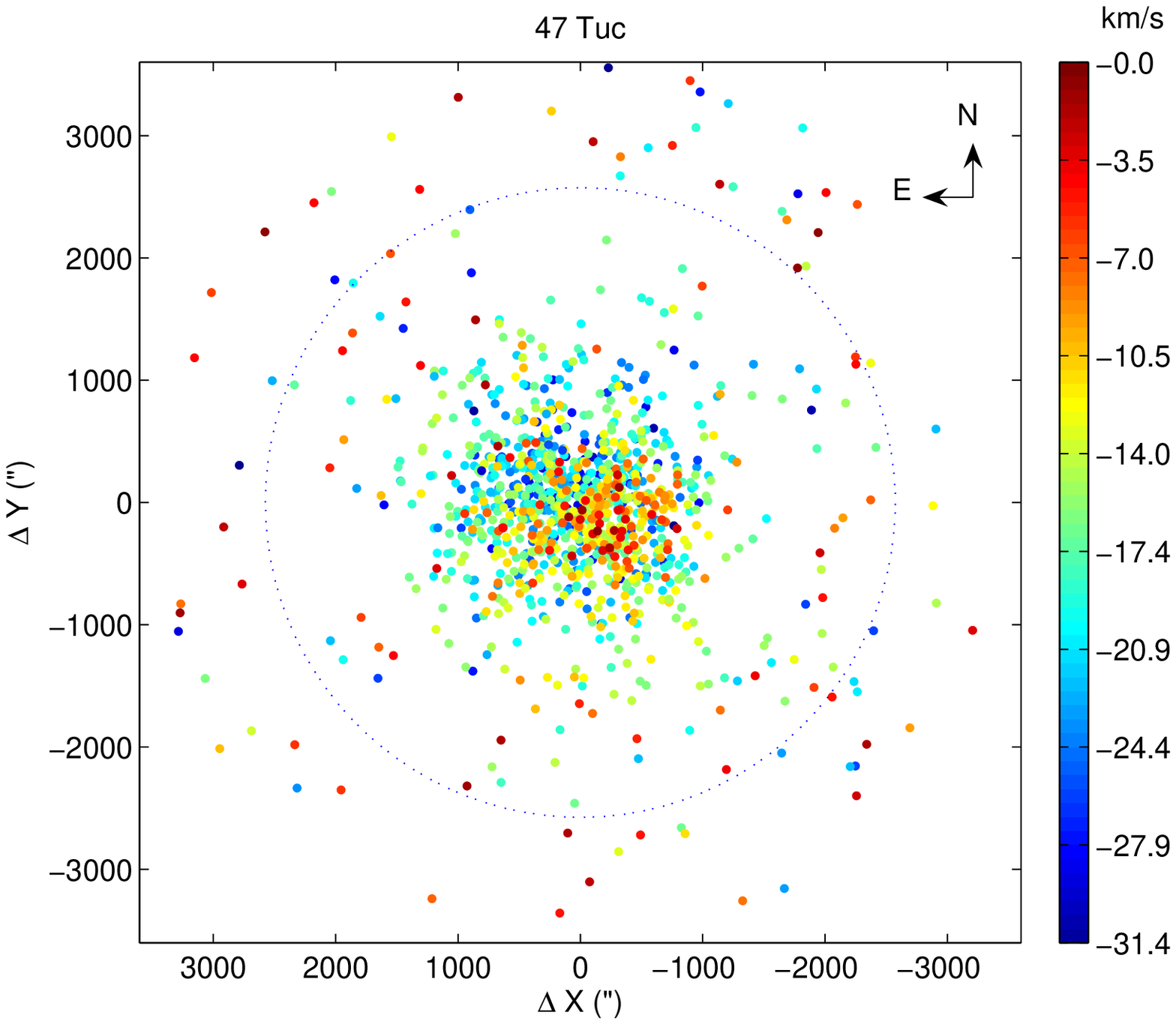}
\includegraphics[width=7cm]{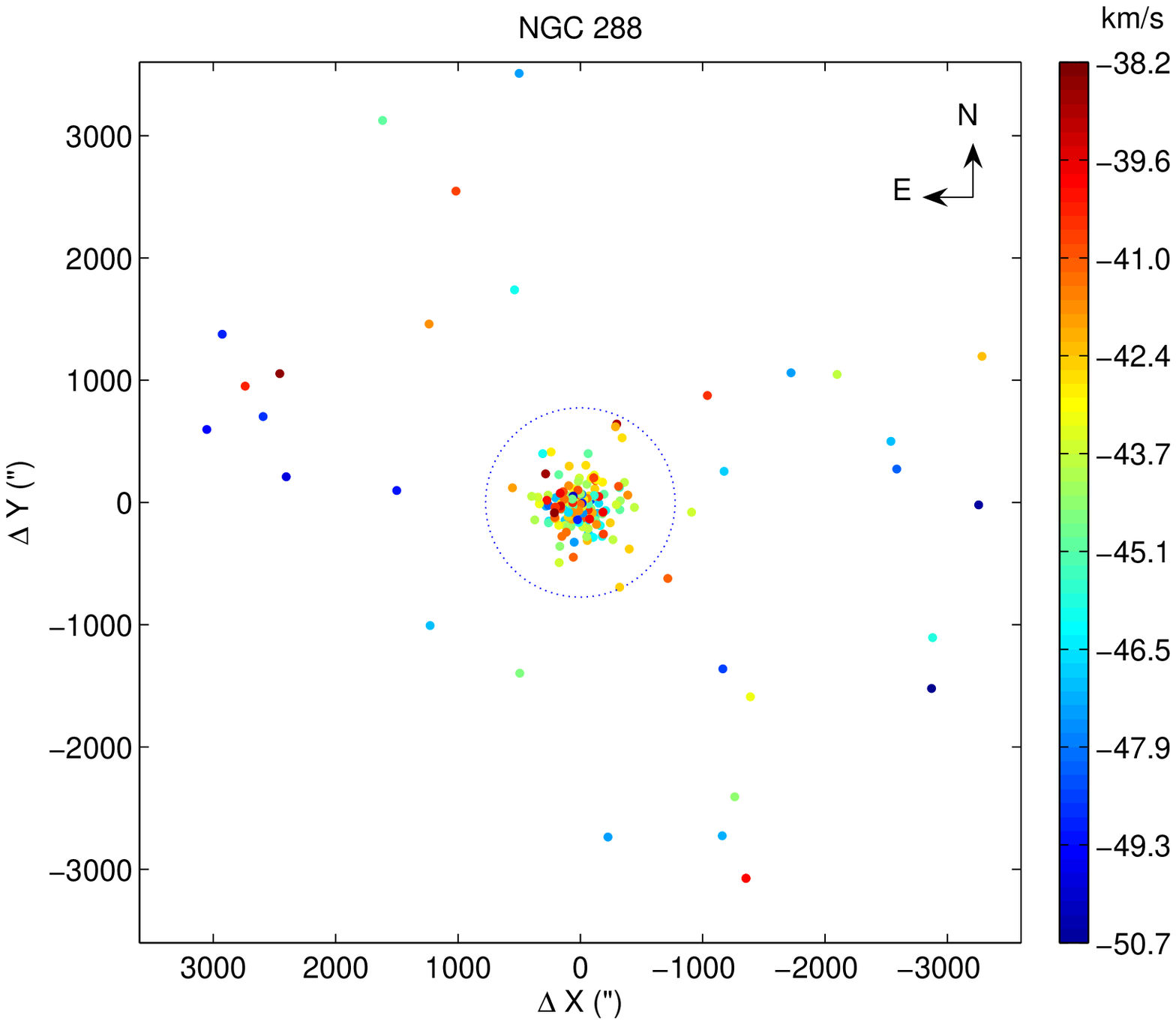}
\includegraphics[width=7cm]{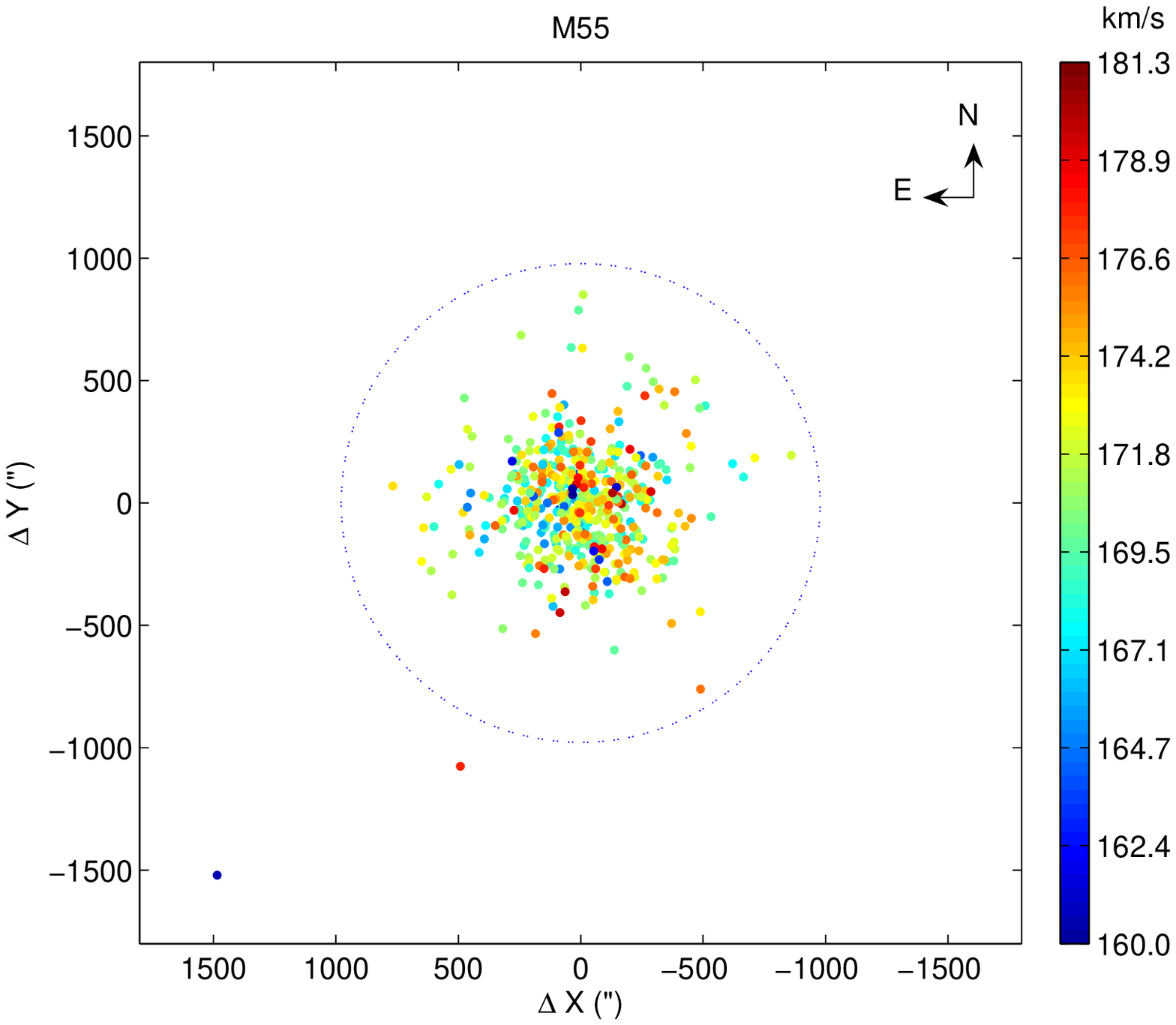}
\includegraphics[width=7cm]{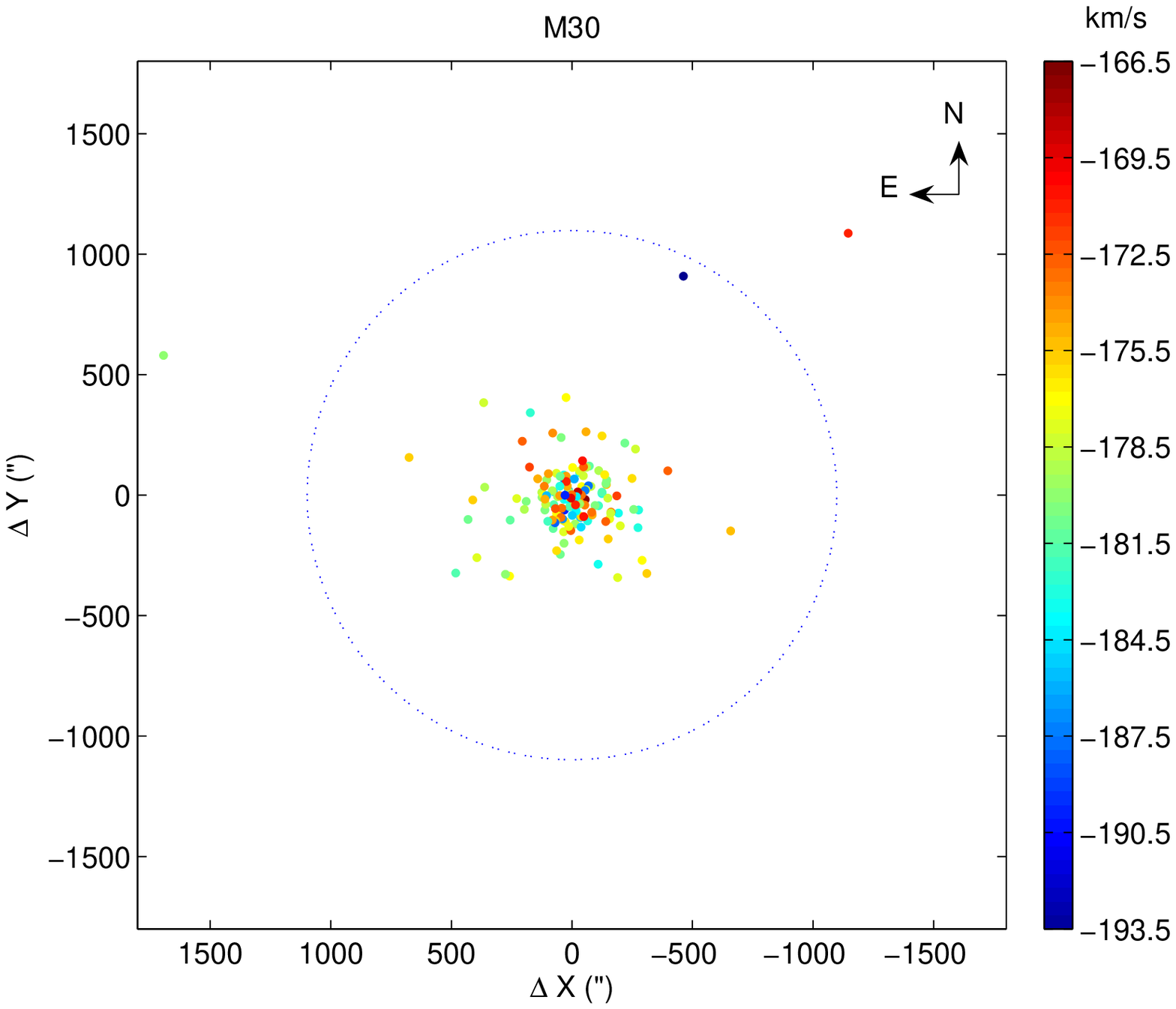}
\caption[]{\label{4gh} Stellar positions for the other four clusters, color coded by the
radial velocity. Note the NW-SW and E-W asymmetry in 47~Tuc and M55, respectively.} 
\end{center}
\end{figure*}

Figs.\ \ref{m12}-\ref{4gh} show the positions of the stars in the core samples, color
coded by the individual $\Sigma{\rm EW}$ values (Fig.\ \ref{m12}) and RVs (Fig.\
\ref{4gh}). For 47~Tuc, NGC~288 and M12 we show the full two-degree fields of view, while
M55 and M30 are zoomed into the central one-degree field. The only evidence for
non-isotropic distribution is found for NGC~288, where we see a NE-SW elongation that
agrees with the reported tidal tails in the literature (Grillmair et al. 1995; Leon et al.
2000). However, the RVs outside the tidal radius do not show a coherent gradient in any
direction, although stars in leading and trailing tails are expected to have systematic
velocity shifts with opposite signs of a few km/s (Dehnen et al. 2004). On the other
hand, there is a slight excess of stars outside the tidal radius, represented by the
difference $N^\prime_{r_{\rm t}}-N_{\rm hist.}=5$. Being positive only for NGC~288, this
excess may support the existence of the reported tidal tails. To find a more conclusive result,
we will use spectroscopic metallicities to secure membership identification for
extra-tidal stars in a future study. For M12, the distribution of $\Sigma{\rm EW}$ across
the field marks the cluster clearly confined to the tidal radius.

Lauchner et al. (2006) discussed the relation between core concentration and tidal
tails, showing that clusters with lower concentration parameter $c$ are more likely to
have significant extra-tidal debris, and that the cluster concentration is a good
predictor of the presence of significant tidal tails. In our sample, M55 and NGC~288 
have very similarly low concentration than clusters with detected tidal tails 
(Pal~5, NGC~5466, NGC~5053), and yet neither of them has produced a positive detection. 

To characterize the limits of our non-detections, we compared the cumulative
$K$-magnitude distributions of the observed stars with the whole two-degree 2MASS fields
for M55,  M30 and NGC~288. For M30, we observed 87\% of all stars in the field brighter
than 13.40 mag, which is 4.7 mag fainter than the tip of the AGB. For NGC~288, 82\% of
stars brighter than 13.55 mag were observed (the tip of the AGB plus 5.0 mag). This 
means we can safely exclude the presence of tidal debris that contains 1\% or more of
the red giant branch stars in the clusters. For M55, the numbers show  that 50\% of all
stars brighter than 12.9 mag that match the
color-magnitude diagram of the cluster were observed. With almost 500 members
captured, 1\% debris that contains red giant branch stars would have produced two or
three clear detections. This can be firmly excluded considering the large $\Sigma{\rm
EW}$ values of the seven extra-tidal stars around M55 (5.5--7.7\AA). We therefore
conclude that the concentration parameter has a much weaker predictive power than
suggested by Lauchner et al. (2006). 

To further search for any putative tidal debris, a matched-filter analysis  was
undertaken for each of the GC systems (Rockosi et
al. 2002). Stars were  selected in the range  $17.5\ge J  \ge8.5$ and  $-0.5\leq 
(J-K)\leq2.2$  and, for  each  GC,  a background colour-magnitude
distribution  was constructed by considering  stars in an annulus of $35\leq r\leq 58$
arcmins.  The colour-magnitude  distribution was  also  constructed  using  stars 
within the  central  3.5  arcmins.  The distributions  were   binned  and  smoothed  to 
make   them  continuous,  and subtracting a  scaled outer distribution  from the inner 
distribution  clearly displayed the colour-magnitude sequence of the cluster itself. The
cluster  and background distributions  were normalized and the  matched-filter analysis 
was performed  on a  grid of  $128\times128$ in  a box  of 40  arcmins on  a  side,
centred on  each cluster. Each point of  the grid represented the  centre  of a box of
$1\times1$ arcmin for an  initial analysis, and $5\times5$ arcmin  for a secondary
analysis, with the  matched-filter analysis revealing what  number of the  stars in 
that  region probably  belong  to the  cluster. However,  while clearly revealing  the
main  body of each  cluster, the analysis  uncovered no significant  substructure that
could  be interpreted  as coherent tidal  debris.

\section{Future work}

Our spectra form an excellent database for studying the structure and evolution of the
clusters. Fig.\ \ref{4gh} shows several important features that will help us to model
their internal dynamics. Firstly, two clusters, 47~Tuc and M55, are characterized by
large-scale asymmetry in velocity distribution that can be interpreted in terms of
systemic rotation. A preliminary analysis of stellar velocities as function of
position angle inferred a rotational velocity of 6$\pm$1 km/s and 3$\pm$1 km/s
for 47~Tuc and M55, respectively, of which the 47~Tuc result is in excellent agreement 
with the 6.5 km/s obtained by Meylan \& Mayor (1986). The influence of rotation on
dynamical evolution may be significant, thus measuring it gives important clues on
realistic models (Combes et al. 1999). Our data are the first to show clear signs of
rotation for M55. Secondly, all five clusters are characterized by large velocity
dispersions in the center (blue and red dots sitting next to each other) that gradually
goes down at the edge (green belts of stars halfway to the tidal radii in Fig.\
\ref{4gh}). Thirdly, we do not see any evidence for increasing dispersion at large radii
that would indicate strong tidal heating by the external gravitational field (Drukier et
al. 1998), the presence of surrounding dark halo (Carraro \& Lia 2000) or a breakdown of
the Newtonian dynamics in the weak acceleration regime (Scarpa et al. 2007). We plan to
test these theories in a future investigation. 

\acknowledgments

This work has been supported by a University of Sydney Research Fellowship, the Hungarian
E\"otv\"os Fellowship and the OTKA Grant \#T042509. Support for program number
HST-HF-01170.01-A to G.\'A.B. was provided by NASA through a Hubble Fellowship grant from
the STScI, which is operated by the AURA, Inc., under NASA contract  NAS5-26555.
G.~\'A.~B.~also wishes to thank useful discussions to A.~P\'al and B.~Cs\'ak. We are
grateful to the staff of the Anglo-Australian Observatory for the superb assistance
during the observations.

\clearpage

\end{document}